\documentclass[preprint2]{aastex63}

\usepackage{graphicx,epsfig,fancyhdr,psfig,rotating,amsmath,epsf,txfonts,natbib,epstopdf,multirow,booktabs,color}

\submitjournal{ApJ}

\shorttitle{Type II Radio Burst}
\shortauthors{Hou et al.}

\def \kms {{\rm km\;s$^{-1}$}}
\def \mhzs {{\rm MHz\;s$^{-1}$}}
\def \arcsec {$^{''}$}
\def \ha {H$\alpha$}

\def \nevii {Ne\,{\sc vii}}
\graphicspath{{./}}

\begin{document}
\title{A Type II Radio Burst Driven by a Blowout Jet on the Sun}

\correspondingauthor{Hui Tian}
\email{huitian@pku.edu.cn}

\author{Zhenyong Hou}
\affiliation{School of Earth and Space Sciences, Peking University, Beijing, 100871, China}
\affiliation{Key Laboratory of Solar Activity and Space Weather, National Space Science Center, Chinese Academy of Sciences, Beijing 100190, China}

\author{Hui Tian}
\affiliation{School of Earth and Space Sciences, Peking University, Beijing, 100871, China}
\affiliation{Key Laboratory of Solar Activity and Space Weather, National Space Science Center, Chinese Academy of Sciences, Beijing 100190, China}
\affiliation{National Astronomical Observatories, Chinese Academy of Sciences, Beijing, 100011, China}

\author{Wei Su}
\affiliation{TianQin Research Center for Gravitational Physics $\&$ School of Physics and Astronomy, Sun Yat-sen University (Zhuhai Campus), Zhuhai 519082,  People's Republic of China}

\author{Maria S. Madjarska}
\affiliation{Max Planck Institute for Solar System Research, Justus-von-Liebig-Weg 3, 37077, G\"ottingen, Germany}
\affiliation{Space Research and Technology Institute, Bulgarian Academy of Sciences, Acad. Georgy Bonchev Str., Bl. 1, 1113, Sofia, Bulgaria}

\author{Hechao Chen}
\affiliation{School of Physics and Astronomy, Yunnan University, Kunming 650050, China}

\author{Ruisheng Zheng}
\affiliation{School of Space Science and Physics, Institute of Space Sciences, Shandong University, Weihai, Shandong, 264209, China}

\author{Xianyong Bai}
\affiliation{National Astronomical Observatories, Chinese Academy of Sciences, Beijing, 100011, China}
\affiliation{Key Laboratory of Solar Activity and Space Weather, National Space Science Center, Chinese Academy of Sciences, Beijing 100190, China}
\affiliation{School of Astronomy and Space Science, University of Chinese Academy of Sciences, Beijing 101408, China}

\author{Yuanyong Deng}
\affiliation{National Astronomical Observatories, Chinese Academy of Sciences, Beijing, 100011, China}
\affiliation{Key Laboratory of Solar Activity and Space Weather, National Space Science Center, Chinese Academy of Sciences, Beijing 100190, China}
\affiliation{School of Astronomy and Space Science, University of Chinese Academy of Sciences, Beijing 101408, China}

\begin{abstract}
Type II radio bursts are often associated with coronal shocks that are typically driven by coronal mass ejections (CMEs) from the Sun.
Here, we conduct a case study of a type II radio burst that is associated with a C4.5 class flare and a blowout jet, but without the presence of a CME.
The blowout jet is observed near the solar disk center in the extreme-ultraviolet (EUV) passbands with different characteristic temperatures.
Its evolution involves an initial phase and an ejection phase with a velocity of 560$\pm$87\,\kms.
Ahead of the jet front, an EUV wave propagates at a projected velocity of $\sim$403$\pm$84\,\kms\ in the initial stage.
The moving velocity of the source region of the type II radio burst is estimated to be $\sim$641\,\kms, which corresponds to the shock velocity against the coronal density gradient.
The EUV wave and the type II radio burst are closely related to the ejection of the blowout jet,
  suggesting that both are likely the manifestation of a coronal shock driven by the ejection of the blowout jet.
The type II radio burst likely starts lower than those associated with CMEs.
The combination of the velocities of the radio burst and the EUV wave yields a modified shock velocity at $\sim$757\,\kms.
The Alfv\'en Mach number is in the range of 1.09--1.18, implying that the shock velocity is 10\%--20\% larger than the local Alfv\'en velocity.
\end{abstract}

\section{Introduction}
\label{sec:intro}

Solar type II radio bursts, which are recognized as the hallmarks of coronal shocks induced by solar activities,
  display as bright features slowly descending to lower frequencies with time in radio dynamic spectra \citep[e.g.,][]{1947Natur.160..256P,1953Natur.172..533W}.
The frequency drifts indicate that coronal shocks propagate outwards along a direction of decreasing coronal density.
Furthermore, coronal shocks can take the form of coronal wave-like phenomena, commonly referred to as extreme ultraviolet waves \citep[EUV waves, e.g.,][]{2014SoPh..289.3233L,2016GMS...216..381C}.
Band-splits of radio emission are commonly observed in type II radio bursts, which are generally considered to be a result of the plasma emission from the upstream and downstream shock regions
  \citep{1974IAUS...57..389S,1975ApL....16...23S,2001A&A...377..321V}.
Therefore, band-splits are utilized to deduce key properties of type II bursts or coronal shocks,
  including the density ratio between the upstream and downstream shock regions and the shock Alfv\'en Mach number
  \citep[e.g.,][]{2009ApJ...691L.151L,2011ApJ...738..160M,2012ApJ...744...72G,2014SoPh..289.2123K,2016ApJ...830...70S}.
\cite{2014ApJ...793L..39D} reported a special type II radio burst with a temporal shift between splitting bands.
Based on their findings, they proposed that the band-splitting signals are moderately polarized with left-handed polarized emission stronger than the right-hand one.
It is commonly believed that coronal shocks result from the coronal mass ejection (CME) propagation or expansion
  \citep[e.g.,][]{2004SoPh..225..105C,2008SoPh..253..215V,2009ApJ...691L.151L,2011LRSP....8....1C,2012ApJ...746...13L,2012ApJ...752L..23S,2013ApJ...765..148C,2013ApJ...767...29F,2014ApJ...787...59C,2018ApJ...856...24Y,2020FrASS...7...17M,2022ApJ...931L...8L,2022ApJ...937L..21Z}.

Solar coronal jets are another possible driver of coronal shocks that manifest as type II radio bursts \citep{2001JGR...10625135G}.
They are a type of pervasive and explosive phenomena in the solar atmosphere, which have a morphology characterized by a nearly collimated plasma ejection following footpoint brightening
 \citep[e.g.,][]{2011A&A...526A..19M,2012ApJ...748..106T,2016SSRv..201....1R,2021RSPSA.47700217S,2018ApJ...861..108Z}.
Previous studies have shown that solar jets are closely associated with type III radio bursts
  \citep[e.g.,][]{2008A&A...491..279C,2008ApJ...675L.125N,2009A&A...508.1443B,2011A&A...531L..13I,2012ApJ...754....9G,2013ApJ...769...96C,2016SSRv..201....1R,2022ApJ...926L..39D}.
Type II radio bursts associated with solar jets are rarely reported \citep{2015ApJ...804...88S,2020ApJ...893..115C,2021ApJ...909....2M,2022ApJ...926L..39D,2023arXiv230511545M}.
From limb observations, \citet{2015ApJ...804...88S} reported a type II radio burst accompanied by a solar jet.
They proposed that this radio burst was caused by the expansion of strongly inclined magnetic loops after reconnecting with nearby emerging flux.
Similarly, \cite{2023arXiv230511545M} reported the source region of a type II radio burst ahead of disturbed coronal loops.
\citet{2020ApJ...893..115C} presented a type II burst showing a continuous transition from a stationary to a drifting state, which was accompanied by a solar jet.
They suggested that the radio burst was likely generated by the interaction between the jet-like CME-driven shock and a streamer.
A type II radio burst with a source region located half solar radii above a solar jet and propagating with a velocity of 1000\,\kms\ was reported by \citet{2021ApJ...909....2M},
  while the related solar jet propagated much more slowly with a velocity of 200\,\kms.
In another study, \citet{2022ApJ...926L..39D} presented a solar jet associated with many phenomena, including a narrow jet-like CME, a type II radio burst, and a type III radio burst.
The authors suggested that this type II radio burst was generated at a height of 1.57--1.68 solar radii with a shock velocity of $\sim$286\,\kms.

As mentioned above, the relationship between solar jets and type II radio bursts has not been extensively studied and further research is required to establish their connection.
In this study, we present an investigation into the properties and relationship between a solar jet, an EUV wave, and a type II radio burst.
Specifically, we report a coronal shock that is closely related to the ejection of a solar jet.
We describe the observations in Section\,\ref{sec:obs}, present the analysis results in Section\,\ref{sec:res}, and the discussion in Section\,\ref{sec:dis}.
We summarize our findings in Section\,\ref{sec:sum}.

\section{Observations}
\label{sec:obs}

\begin{figure*}
\centering
\includegraphics[trim=0.0cm 0.3cm 0.0cm 0.0cm,width=0.9\textwidth]{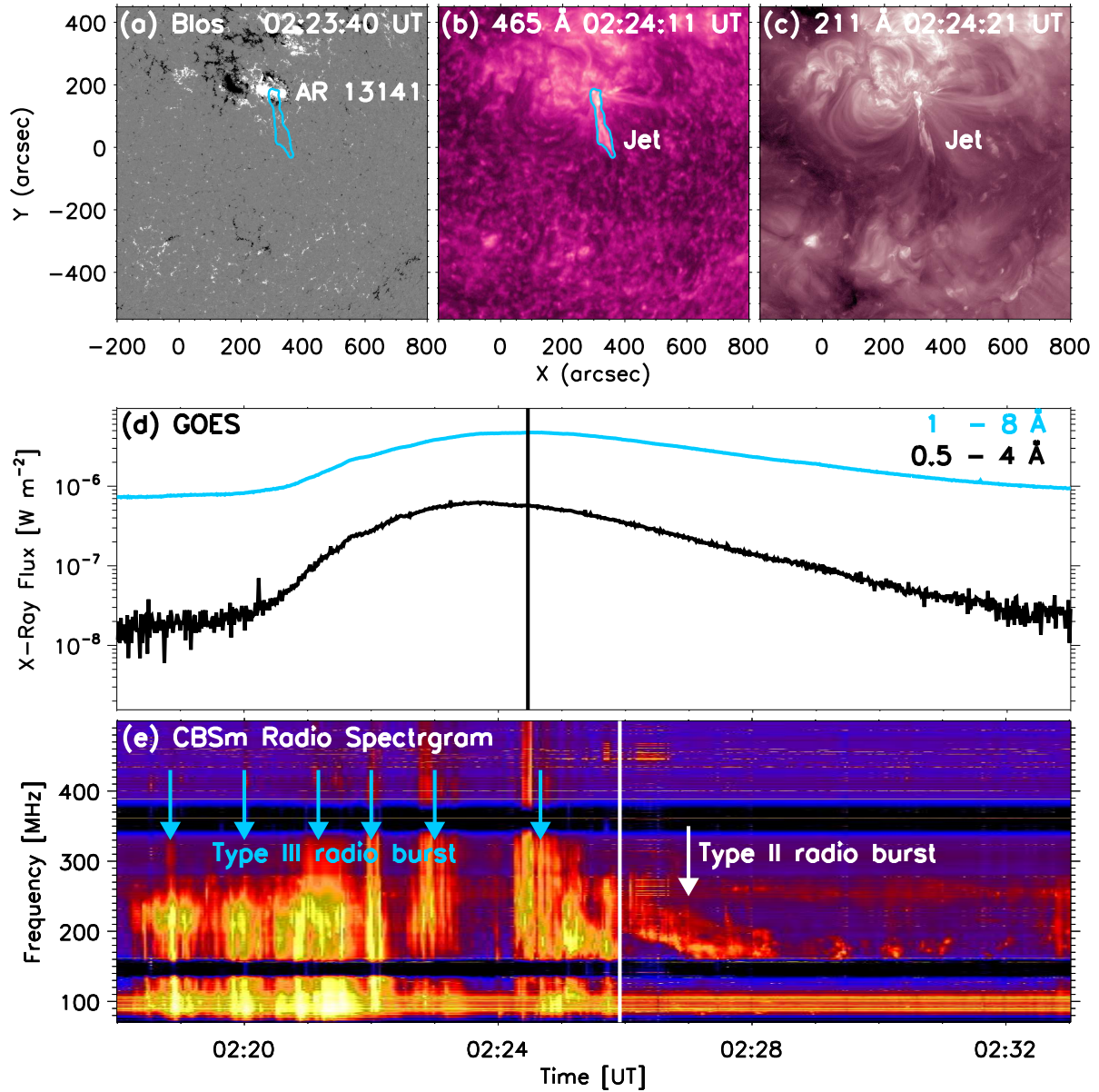}
\caption{Overview of the event from the Earth's view.
(a) HMI LOS magnetogram at 02:23:40 UT, saturated at $\pm$200\,G.
(b) SUTRI 465\,\AA\ image at 02:24:11 UT.
(c) AIA 211\,\AA\ image at 02:24:21 UT.
The cyan contour overplotted in (a) and (b) outlines the solar jet.
(d) Light curves of GOES  X-ray flux at 0.5--4\,\AA\ (black) and 1--8\,\AA\ (blue).
(e) Radio dynamic spectrum on 2022 November 12 observed by CBSm.
The black vertival line in (d) marks the peak time of the X-ray flux at 1--8\,\AA, while the white vertical line in (e) marks the starting time of the type II radio burst.
In (e), the white arrow and the blue arrows indicate the type II radio burst and the type III radio bursts, respectively.}
\label{fig:overview}
\end{figure*}

On November 12, 2022, a type II radio burst was detected by the Chashan Broadband Solar radio spectrograph at meter wavelength (CBSm).
CBSm, a newly constructed instrument located at the Chashan Solar Observatory (CSO), is dedicated to monitoring the fine structures of solar radio bursts in the metricwave range.
CSO is managed by the Institute of Space Sciences of Shandong University.
The radio observation used in this study covers a frequency range of  90--600 MHz, with a frequency resolution of 76.29 kHz and a time cadence of 0.84 ms.

A solar jet associated with the radio burst was located in the NOAA active region (AR) 13141 and detected simultaneously by multiple instruments, including
  the Atmospheric Imaging Assembly \citep[AIA,][]{2012SoPh..275...17L} onboard the Solar Dynamics Observatory \citep[SDO,][]{2012SoPh..275....3P}, 
  the Solar Upper Transition Region Imager \citep[SUTRI,][]{2023RAA....23f5014B} onboard the Space Advanced Technology demonstration satellite (SATech-01),
  the ground-based New Vacuum Solar Telescope \citep[NVST,][]{Liu_2014,2016NewA...49....8X,Yan2020},
  and the Extreme Ultraviolet Imager \citep[EUVI,][]{2008SSRv..136...67H} onboard the Solar Terrestrial Relations Observatory \citep[STEREO,][]{2008SSRv..136....5K}.
Images from all of the AIA EUV passbands were used to analyze the solar jet and its associated phenomena.
They have a pixel size of 0.6\arcsec\ and a cadence of 12 s.
SUTRI captures the full-disk transition region images in the \nevii\ 465\,\AA\ line,
  which is formed at a temperature regime of $\sim$0.5 MK \citep{2017RAA....17..110T} in the solar atmosphere between the chromosphere and the corona (the so-called transition region).
SUTRI image has a pixel size of 1.2\arcsec\ and a cadence of 30 s.
The evolution of the solar jet was also examined using the NVST chromospheric images taken in the \ha-center with a cadence of $\sim$73 s and a pixel size of 0.165\arcsec.
The STEREO-Ahead satellite, was located about 15.1$^{\circ}$ west of Earth, recorded the solar jet using the EUVI 195\,\AA\ passband.
The EUVI 195\,\AA\ image has a pixel size of 1.58\arcsec\ and a cadence of 2.5 minutes.
Furthermore, line-of-sight (LOS) magnetic field data obtained by the Helioseismic and Magnetic Imager \citep[HMI,][]{Schou2012hmi} onboard SDO
  were used to visualize the magnetic field structures around the footpoint region of the solar jet.
The pixel size of the magnetogram is 0.5\arcsec, and the cadence is 45 s.

The SUTRI 465 \,\AA\ and AIA 304 \,\AA\ images were aligned using a linear Pearson correlation analysis,
  while the NVST \ha-center images and the AIA 1600\,\AA\ images were aligned using the sunspot and its surrounding plage region within AR 13141 as reference features.

\begin{figure*}
\centering
\includegraphics[trim=0.0cm 0.9cm 0.0cm 0.0cm,width=0.8\textwidth]{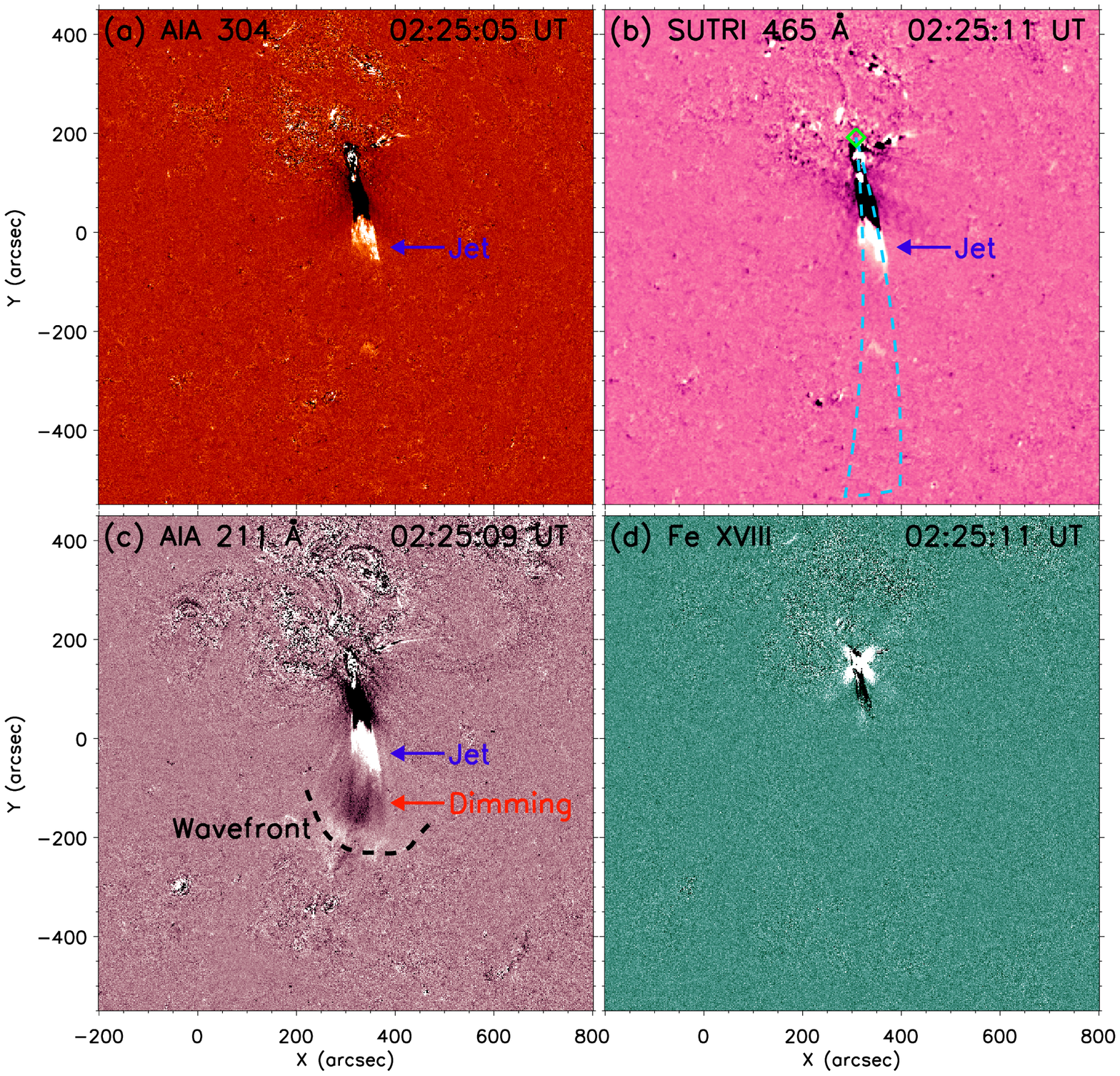}
\caption{Running-difference images of AIA 304\,\AA, SUTRI 465\,\AA, AIA 211\,\AA, and Fe\,{\sc xviii} 93.9\,\AA\ around 02:25 UT showing the solar jet and EUV wave.
The blue arrows in (a)--(d) indicate the solar jet.
In (b), the cyan line represents a sector cut used to obtain the time-distance diagrams shown in Fig.\,\ref{fig:stimg},
  while the green diamond indicates the starting position.
In (c), the black dashed line outlines the wavefront, and the red arrow indicates the dimming region between the wavefront and the solar jet.
An animation of this figure is available, showing the evolution of the solar jet and EUV wave.
It covers a duration of $\sim$20 minutes from 02:15 to 02:35 UT on 2022 November 12.}
\label{fig:maps}
\end{figure*}

\begin{figure*}
\centering
\includegraphics[trim=0.0cm 0.7cm 0.0cm 0.0cm,width=0.85\textwidth]{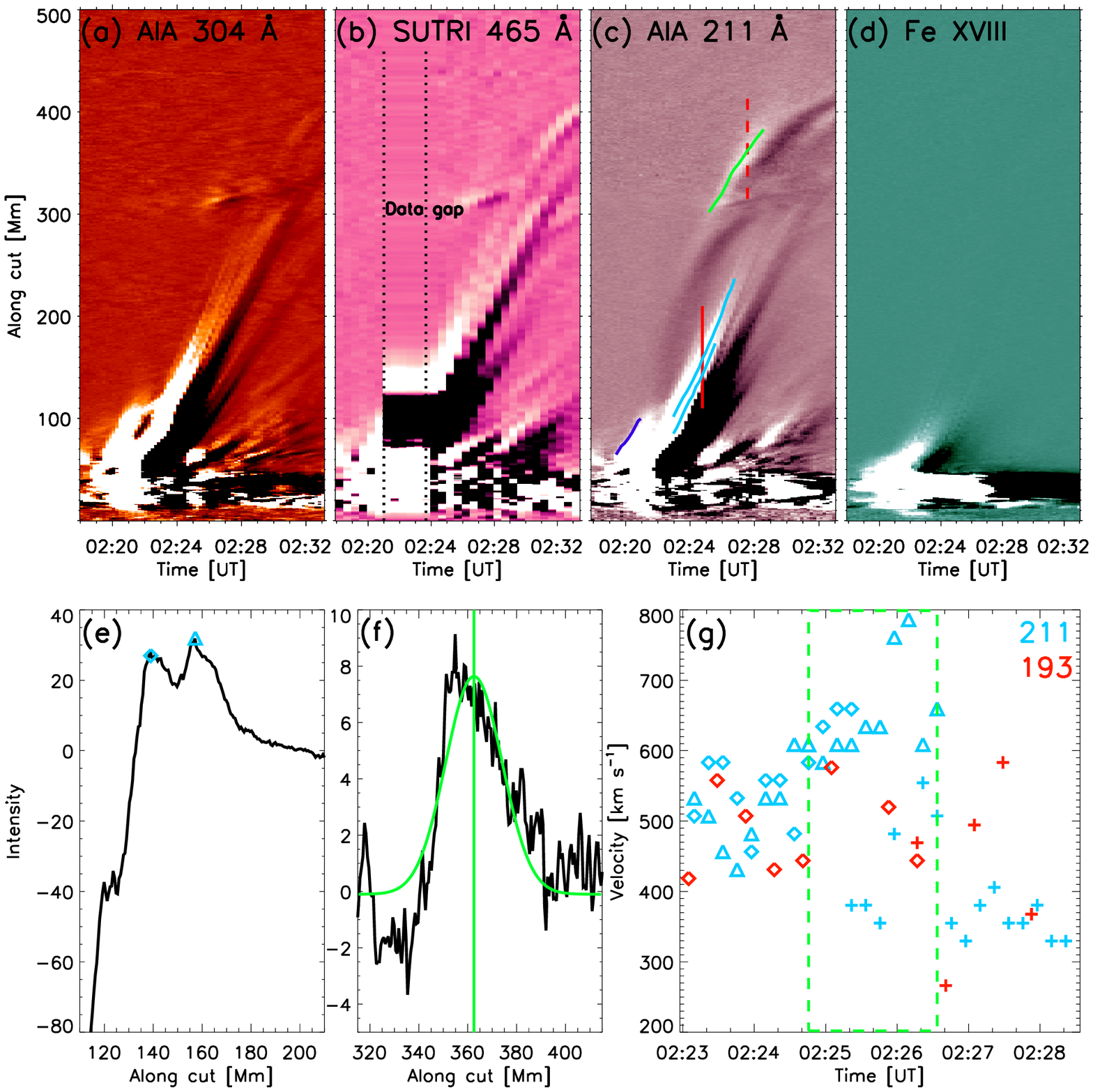}
\caption{Time-distance diagrams showing the propagations of the solar jet and EUV wave,
  as observed in the AIA 304\,\AA, SUTRI 465\,\AA, AIA 211\,\AA, and Fe\,{\sc xviii} 93.9\,\AA\ running-difference images.
In (b), two black dotted lines indicate a data gap.
In (c), the blue, cyan, and green lines represent the propagation of the jet front and wavefront.
(e) and (f) Intensity variations (black lines) along the red solid line and red dashed line in (c).
In (e), two peaks are visible and shown as a diamond and a triangle symbol.
In (f), the green line results from a single Gaussian fit and the vertical green line is the center of the Gaussian.
(g) Velocities of the solar jet (diamond and triangle symbols) and EUV wave (plus symbols) obtained from AIA 211\,\AA\ (cyan) and 193\,\AA\ (red) images.
The green dashed box shows the accelerated propagations of the solar jet and the EUV wave.
}
\label{fig:stimg}
\end{figure*}

\section{Results}
\label{sec:res}

Figure\,\ref{fig:overview} presents an overview of the event in AR\,13141, located near the center of the solar disk.
A solar jet was observed in the south of AR\,13141, indicated by the cyan contour in Figure\,\ref{fig:overview}(a)--(b).
The jet appeared around 02:18:45\,UT and propagated southward, as observed in the AIA 211\,\AA\ and SUTRI 465\,\AA\ images.
As shown in the HMI LOS magnetogram in Figure\,\ref{fig:overview}(a), its footpoint is situated in a region with opposite magnetic polarities, including a dominant positive polarity and several minor negative polarities.

Additionally, a C4.5 class flare was detected by GOES, shown in Figure\,\ref{fig:overview}(d), and lasted for approximately 12 minutes, with onset and peak at 02:20:00 UT and 02:24:27 UT, respectively.
The start of the flare was consistent with the occurrence of the solar jet.
We examined the coronagraph images taken by the Large Angle Spectroscopic Coronagraphonboard \citep{1995SoPh..162..357B} onboard the Solar and Heliospheric Observatory \citep[SOHO,][]{1998GeoRL..25.2465T} and
  the Sun-Earth Connection Coronal and Heliospheric Investigation inner coronagraph COR2 \citep{2008SSRv..136...67H} onboard STEREO-Ahead using the Helioviewer website\footnote{https://helioviewer.org} and the SOHO LASCO CME catalog\footnote{https://cdaw.gsfc.nasa.gov/CME\_list},
  and found that no CME was related to the solar jet.

Figure\,\ref{fig:overview}(e) displays the radio dynamic spectrum taken by CBSm from 02:18 to 02:33 UT.
The CBSm radio dynamic spectrum reveals a clear type II radio burst that undergoes a slow frequency drifting from 02:25:55 UT (indicated by the white vertical line in Figure\,\ref{fig:overview}(e)) to 02:28:30 UT.
This radio burst began about 88 s after the peak time of the GOES X-ray flux, as indicated in Figure\,\ref{fig:overview}(d) and (e).
We examined the EUV images and did not identify any other noteworthy eruptions in the solar atmosphere during the same time period.
Therefore, we confirm that this type II radio burst is associated with the solar jet and the C4.5 class flare.
Moreover, several type III radio bursts appeared in the time period from 02:18 UT to 02:26 UT, which might be associated with the generation of the solar jet.

\subsection{Solar jet and EUV wave from the Earth's perspective}
\label{subsec:earth}

From the Earth's perspective, the solar jet was visible in the EUV passbands of SUTRI and AIA.
Figure\,\ref{fig:maps} and its corresponding animation illustrate the region of interest in the running-difference images with a 2-minute time separation of the AIA 304\,\AA\ \citep[log T(K) $\sim$ 4.7,][]{2003A&A...400..737A},
  SUTRI 465\,\AA\ (log T(K) $\sim$ 5.7), AIA 211\,\AA\ (dominated by Fe\,{\sc xiv} line (log T$_{max}$ (K) $\sim$ 6.3) with cooler emission contribution) and AIA 94\,\AA\ passbands.
The AIA 94\,\AA\ channel is dominated by Fe\,{\sc xviii} 93.9\,\AA\ (log T(K) $\sim$ 6.8) and cooler emission that was removed by applying the expression derived by \cite{2013A&A...558A..73D}.
Thus, we will refer to it hereafter as AIA Fe XVIII. 
The solar jet is prominently visible in the cool and warm passbands, such as the AIA 304\,\AA, SUTRI 465\,\AA, AIA 211\,\AA,
  but very faint in the high-temperature passband, i.e., Fe\,{\sc xviii} 93.9\,\AA\ passband.
It propagates southward along a closed loop system starting from the edge of AR\,13141 and reaches the far footpoint region of this loop system (see the corresponding animation of Figure\,\ref{fig:maps}).
The far foopoint region includes several minor negative magnetic concentrations
  situated approximately at X$\sim$250\arcsec\ and Y$\sim$-300\arcsec\ (see the Figure\,\ref{fig:overview}(a) and Figure\,\ref{fig:maps}).
A wave-like structure, undergoing a southward propagation, is visible only to the south of the jet, and its outer edge is highlighted by the black dashed line in Figure\,\ref{fig:maps}(c).
The AIA 211\,\AA\ passband reveals an apparent signature of the propagating wave-like structure, while it is weak in AIA 193\,\AA\ and indiscernible in the other EUV passbands. 
Additionally, during the evolution of these events, a dimming region appears between the solar jet and the wave-like structure in AIA 211\,\AA.

We have defined a sector cut that starts from the footpoint region of the solar jet and run nearly parallel to the propagations of the solar jet and wave-like structure.
This cut allows us to obtain time-distance diagrams for the running-difference images of the AIA 304\,\AA, SUTRI 465\,\AA, AIA 211\,\AA, and Fe\,{\sc xviii} passbands, as shown in Figure\,\ref{fig:stimg}(a)--(d).

The EUV time-distance diagrams in Figure\,\ref{fig:stimg}(a)--(d) illustrate the propagation of the solar jet that includes two phases:
  an initial phase from $\sim$02:18 UT with a low velocity and an ejection phase from $\sim$02:22 UT with a high velocity.
These two phases are further described in Section\,\ref{subsec:jetfoot}.
For the initial phase of the solar jet, its propagation is determined visually and shown as the blue line in Figure\,\ref{fig:stimg}(c),
  because the footpoint region has complex intensity variations.
Figure\,\ref{fig:stimg}(e) displays the intensity variation of the jet front along the red solid line in Figure\,\ref{fig:stimg}(c).
It is hard to use a single Gaussian fit to confirm the location of the jet front due to two intensity peaks therein.
These two peaks are visible with time and can be determined visually (shown as a diamond and a triangle symbol in Figure\,\ref{fig:stimg}(e)).
The arrays of the first peak and the second peak are outlined with the lower cyan line and the upper cyan line in Figure\,\ref{fig:stimg}(c), respectively, which can both represent the propagation of the jet front.
Figure\,\ref{fig:stimg}(f) displays the intensity variation of the wavefront (black line) along the red dashed line in Figure\,\ref{fig:stimg}(c).
To determine the location of the wavefront, we applied a single Gaussian fit (green line).
In Figure\,\ref{fig:stimg}(f), the vertical green line is the center of the Gaussian and marks a location of the wavefront.
Intensity variations of the wavefront with higher values than surrounding positions are used to determine the propagation of the wavefront, which is shown as the green line in Figure\,\ref{fig:stimg}(c).

The propagation tracks are similar in AIA 304\,\AA, SUTRI 465\,\AA\ and AIA 211\,\AA, while the Fe\,{\sc xviii} passband shows a shorter propagation distance.
Notably, the AIA 211\,\AA\ time-distance diagram clearly shows a bright stripe representing the propagation of the wave-like structure as indicated by the green line in Figure\,\ref{fig:stimg}(c).
Through careful analysis, we found that the wave-like structure appeared at 02:25:09--02:28:33 UT in AIA 211\,\AA\ and at 02:25:52--02:28:16 UT in AIA 193\,\AA.
The wave-like structure starts about three minutes after the start of the ejection phase of the solar jet and less than one minute ahead of the start of the type II radio burst.
Furthermore, the dimming region between the solar jet and the wave-like structure is also visible in the time-distance diagram of AIA 211\,\AA.

Considering the wave-like structure appearing only in AIA 211\,\AA\ and 193\,\AA, we determined the wave-like structure velocity using the AIA 193\,\AA\ and 211\,\AA\ images.
The velocities during the initial and ejection phases of the solar jet were determined also from the images of these two EUV passbands.
The initial velocity of the solar jet is almost constant and estimated to be 370$\pm$19 \kms\ (averaged from 384 \kms\ in AIA 211\,\AA\ and 357 \kms\ in AIA 193\,\AA) by applying a linear fit.
The projected velocities of the solar jet and the wave-like structure were calculated with time by applying linear fits to every three propagation locations.
The time cadence between two propagation locations is 12--24 s.
Figure\,\ref{fig:stimg} (g) displays the derived velocities in AIA 211\,\AA\ and 193\,\AA.
The propagation velocity of the solar jet is in the range of 431--786 \kms\ in AIA 211\,\AA\ and 418--516 \kms\ in AIA 193\,\AA\ with an average value of 560$\pm$87 \kms.
The wave-like structure propagates at a velocity of 403$\pm$84 \kms\ (329--554 \kms\ in AIA 211\,\AA\ and 266--583 \kms\ in AIA 193\,\AA),
  which is accelerated to 554\,\kms\ around 02:26:30 UT.
At the same time, the solar jet undergoes an accelerated propagation (see the green dashed box in Figure\,\ref{fig:stimg}(g)),
  and the type II radio burst can be seen in the radio dynamic spectrum (see Figure\,\ref{fig:overview}(e)). 

The wave-like structure obtained from AIA 193\,\AA\ and 211\,\AA\ is similar to that of the coronal wave-like phenomena observed during solar eruptions
  \citep[i.e., EUV wave,][]{2012ApJ...752L..23S,2014SoPh..289.3233L,2022ApJ...928...98H}.
The EUV waves have often been interpreted as fast-mode MHD shocks or waves driven by CMEs, flares, or other small-scale events
  \citep[e.g.,][]{2012ApJ...753L..29Z,2013ApJ...764...70Z,2014SoPh..289.3233L,2015ApJ...812..173V,2016GMS...216..381C,2021SoPh..296..169Z},
  or a CME-driven compression \citep[non-wave component, e.g.,][]{2002ApJ...572L..99C,2005ApJ...622.1202C,2022ApJ...939L..18S}.
Our observations revealed that the wave-like feature appeared slightly after the acceleration of the solar jet and was followed by a dimming region.
At the same time, the type II radio burst also appeared.
The dimming region could be consistent with the rarefaction region in the wake of the EUV wave \citep[e.g.,][]{2015LRSP...12....3W}.
These findings are similar to those of CME-driven EUV waves found in previous studies \citep[e.g.,][]{2015A&A...575A..39M,2022ApJ...928...98H}.
Based on our results, we propose that the wave-like structure is likely an EUV wave, which might be interpreted as a fast-mode MHD shock wave that is shortly driven by the ejection of the solar jet.

\subsection{Solar jet and EUV wave from STEREO's perspective}
\label{subsec:stereo}

\begin{figure*}
\centering
\includegraphics[trim=0.0cm 0.3cm 0.0cm 0.0cm,width=0.8\textwidth]{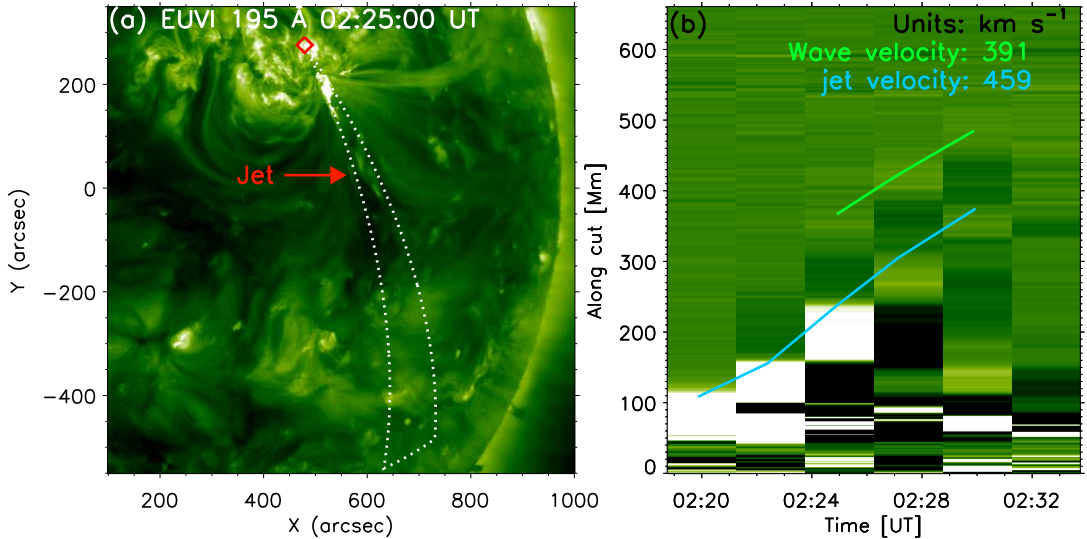}
\caption{(a) EUVI 195\,\AA\ image at 02:25:00 UT.
The white dotted line represents a sector cut used to obtain the time-distance diagram shown in (b), and the red diamond indicates the starting position.
(b) Time-distance diagram showing the propagations of the solar jet (cyan line) and EUV wave (green line), as obtained from the EUVI 195\,\AA\ running-difference images.
An animation of this figure is available, showing the evolution of the solar jet and EUV wave.
It covers a duration of $\sim$20 minutes from 02:15 to 02:35 UT on 2022 November 12.}
\label{fig:stereo}
\end{figure*}

The solar jet and EUV wave were also observed in the EUVI 195\,\AA\ passband from STEREO's perspective,
  which is approximately 15.1$^{\circ}$ to the west of the Sun-Earth line, as shown in Figure\,\ref{fig:stereo} and its accompanying animation.
Figure\,\ref{fig:stereo} and the accompanying animation reveal the southward propagation of the solar jet and the EUV wave,
  along which we defined a sector cut marked by a white dotted line in Figure\,\ref{fig:stereo}(a) to obtain the time-distance diagram.
The time-distance diagram obtained from the running-difference images of the EUVI 195\,\AA\ passband, is shown in Figure\,\ref{fig:stereo}(b).
Although the time cadence was low (2.5 minutes), the propagation tracks of the solar jet and EUV wave are visible in the EUVI 195\,\AA\ time-distance diagram,
  as marked by the cyan and green lines in Figure\,\ref{fig:stereo}(b), respectively.
However, the ejection phase of the solar jet is difficult to differentiate from its initial phase seen in the EUV images taken by SUTRI and AIA.
The dimming region between the solar jet and the EUV wave is well visible.

We also quantified the propagation velocities of the solar jet and EUV wave in the STEREO's perspective.
The velocities of the solar jet and EUV wave are $\sim$459\,\kms\ and $\sim$391\,\kms, respectively, and are shown in Figure\,\ref{fig:stereo}(b) in the same colors as the propagation tracks.
These values are lower than those obtained from the images of the SUTRI and AIA EUV passbands,
  which might be attributed to the different observing perspective and the low temporal resolution.
There is a moving structure in the coronagraph images taken by COR2.
However, this structure appears not to be associated with the jet if we use the obtained jet velocity to predict its trajectory in the field of view of COR2.
Thus, the majority of the jet reaches the other footpoint region of the closed loop system and may not develop into a CME.

\subsection{Coronal shock from the CBSm radio dynamic spectrum}
\label{subsec:shockv}

\begin{figure*}
\centering
\includegraphics[trim=0.0cm 0.2cm 0.0cm 0.0cm,width=0.8\textwidth]{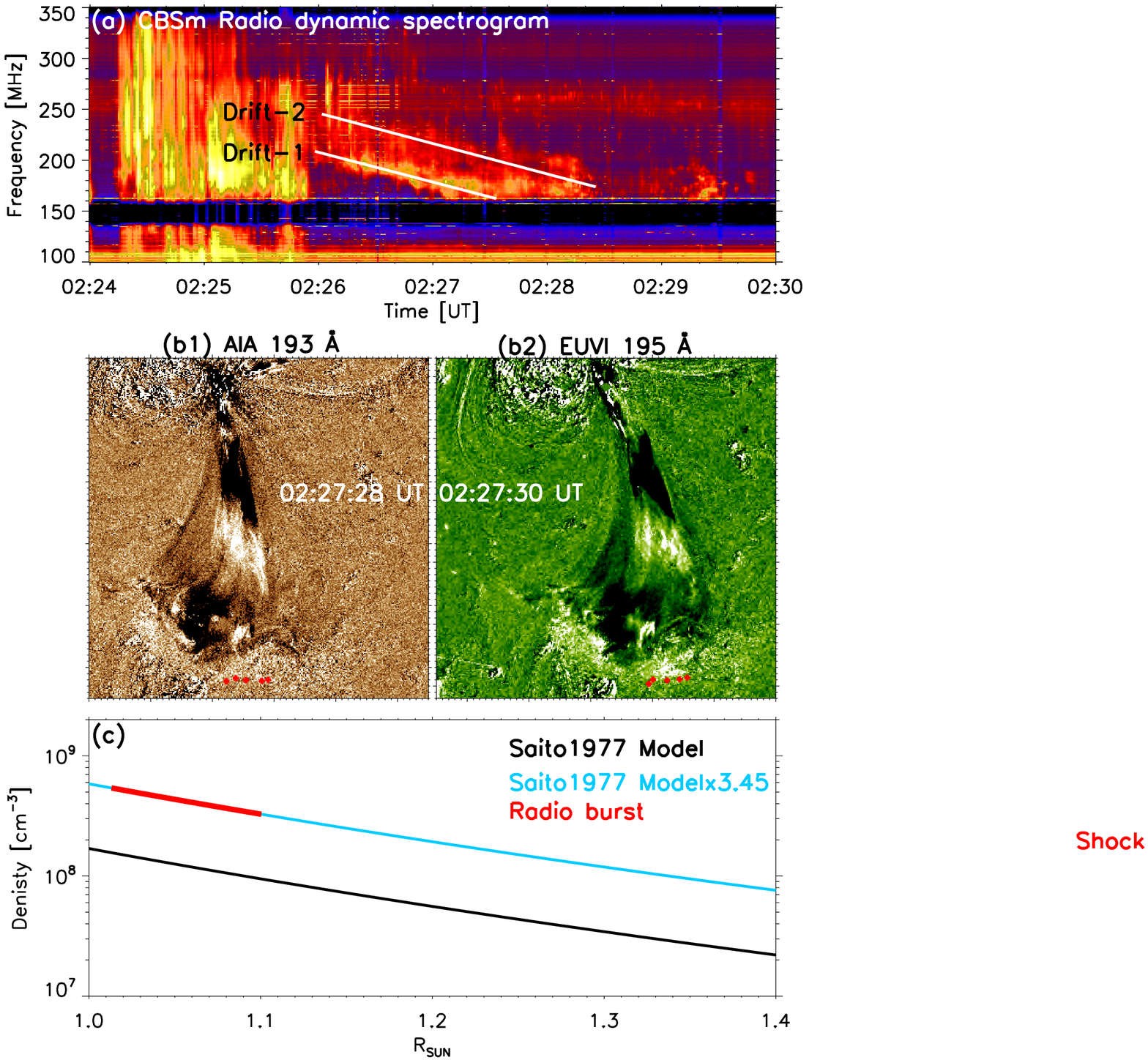}
\caption{(a) Zoomed-in radio dynamic spectrum showing the details of the type II radio burst.
Two white lines (Drift-1 and Drift-2) show the frequency drifts for the upstream and downstream of the shock during the type II radio burst, respectively.
(b1) and (b2) Triangulations of the wavefronts in the images of AIA 193\,\AA\ and EUVI 195\,\AA\ around 02:27:30 UT.
The red symbols are selected to indicate the locations of the wavefronts.
(c) Coronal density as a function of height in solar radii.
The black and blue solid lines represent the coronal density from the model of \citet{1977SoPh...55..121S} and the revised one multiplied by a coefficient of 3.4, respectively.
The red thick line represents the relationship between the height and the coronal density of the shock.}
\label{fig:shocksccv}
\end{figure*}

As previously mentioned, a type II radio burst was detected 46 seconds after the initiation of the EUV wave.
The radio dynamic spectrum taken by CSO/CBSm reveals details of the type II radio burst as depicted in Figure\,\ref{fig:shocksccv}(a).
Generally, the emission from the upstream and downstream shock regions results in a ``band-split" pattern with a similar frequency drifting rate \citep{1975ApL....16...23S}.
For the upstream component, the radio burst commences at a frequency of 208 MHz, while the downstream component starts at 245 MHz.
Two white lines (Drift-1 and Drift-2) in Figure\,\ref{fig:shocksccv}(a) mark the frequency drifts of the upstream and downstream components.
The average drift rates were determined to be $-$0.298\,\mhzs\ and $-$0.304\,\mhzs\ for the upstream and downstream components, respectively.

The frequency drift observed during a Type II radio burst usually corresponds to the density variation of the outward propagating shock.
In this study, we applied a coronal density model obtained from the white-light coronagraph data near the solar cycle minimum, as presented in \citet{1977SoPh...55..121S}, to determine the shock velocity.
The density model is generally constrained by a coefficient multiplication \citep[e.g.,][]{2012ApJ...753...21F,2015ApJ...804...88S}.

First, we used stereoscopic observations to determine the wavefront height around 02:27:30 UT and considered it as the height of the shock source region during that time.
The EUV wave was observed simultaneously by SDO/AIA and STEREO/EUVI in dual perspectives.
Since the AIA 193\,\AA\ passband has a temperature response similar to that of the EUVI 195\,\AA\ passband, they can reveal the same signature of the EUV wave.
This allows us to derive the wavefront height from the stereoscopic observations through the \textsl{Solar Software} procedure scc$\_$measure.pro.
Upon reviewing the observations by SDO/AIA and STEREO/EUVI, we determined that the wavefront was observed simultaneously by these two instruments only at $\sim$02:27:30 UT.
We selected five points, as shown in Figure\,\ref{fig:shocksccv}(b1) and (b2),
  marked by the red symbols along the outer edges of the wavefront not affected by other structures to determine the height of the wavefront.
The derived heights of the wavefront range from 52 to 74 Mm, with an average of 67$\pm$9 Mm.
We then consider the average 67 Mm as the wavefront height at $\sim$02:27:30 UT.

The coronal density as a function of the distance model proposed by \citet{1977SoPh...55..121S} can be expressed in the form
\begin{equation}
N_{e}(r) = c_{1}r^{d_{1}} + c_{2}r^{d_{2}},
\end{equation}
where $r$ is given in units of the solar radii ($R_{S}$), and $c_{1}$ ($\sim 1.36\times10^{6}$), $c_{2}$ ($\sim 1.68\times10^{8}$), $d_{1}$ ($\sim 2.14$), and $d_{2}$ ($\sim 6.13$) are constant coefficients.
The original density variation with height is shown with a black line in Figure\,\ref{fig:shocksccv}(c).
Using the wavefront height of 67 Mm (1.095 $R_{S}$) and the original model, we calculated the corresponding density to be $\sim9.697\times10^{7}$ cm$^{-3}$ ($N_{saito}$).
By relating the frequency of the type II radio burst to the density,
\begin{equation}
f = 8.98\times\sqrt{N_{e}},
\label{fun:fn}
\end{equation}
where $f$ is the upstream frequency in the unit of kHz and $N_{e}$ is the plasma density in the unit of cm$^{-3}$,
  we estimated the coronal density of the shock to be $\sim3.349\times10^{8}$ cm$^{-3}$ ($N_{radio}$) at 02:27:30 UT.

We used the estimated densities $N_{saito}$ and $N_{radio}$ from different observations to revise the original density model from \citet{1977SoPh...55..121S}.
The ratio of $\sim$3.45 between these densities was taken as the coefficient to constrain the original density model.
The revised coronal density model is displayed with a cyan line in Figure\,\ref{fig:shocksccv}(c).
By using Equation\,\ref{fun:fn} and the revised coronal density model, we could derive the variation of coronal density of the shock source region during the type II radio burst
  (refer to the short red line in Figure\,\ref{fig:shocksccv}(c)) through the frequency drift. 
The source region of the radio burst may have a lower height.
We estimated the velocity of the type II radio burst that may have an average value of 641$\pm$64 \kms\ (530--752\,\kms).
This component of the shock velocity likely corresponds to the radial velocity against the coronal density gradient.

In this study, we made a rough estimate on the height of the source region of the radio burst.
During the propagation of a fast-mode magnetoacoustic wave, it may steepen to form a shock when its velocity is larger than the local Alfv\'en velocity.
According to the measurements of the magnetic field from \cite{2011ApJ...730..122W}, \cite{2020ScChE..63.2357Y} and \cite{2020Sci...369..694Y},
  the magnetic field strength in a quiet Sun region may be 1--2 G in the height range of 1.0--1.4 $R_{S}$.
The field strength and the density from the radio dynamic spectrum yield a rough estimate on the local Alfv\'en velocity in the range of 300--760\,\kms.
The velocity of the radio burst is estimated to be $\sim$641\,\kms, indicating that the velocity of the type II radio burst is larger than the local Alfv\'en velocity at a height, where the wave might steepen into a shock.

\subsection{Blowout jet}
\label{subsec:jetfoot}

\begin{figure*}
\centering
\includegraphics[trim=0.0cm 0.9cm 0.0cm 0.0cm,width=0.75\textwidth]{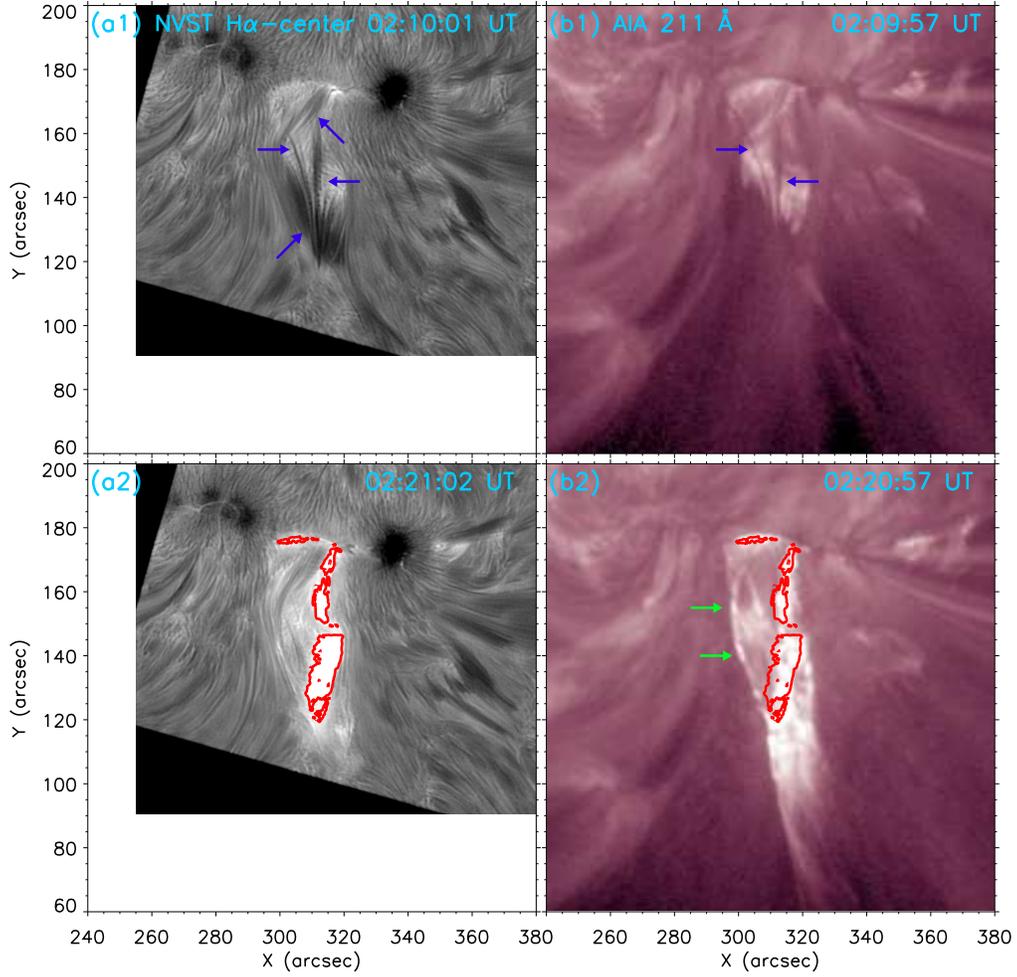}
\caption{Zoomed-in images of the footpoint region of the jet as observed by the NVST \ha-center and AIA 211\,\AA\ images.
(a1)\&(b1) and (a2)\&(b2) are for the times before and during the initial stage of the solar jet, respectively.
In (a1) and (b1), blue arrows mark the signatures of the small filaments.
In (a2) and (b2), the red contours from the NVST \ha-center images mark the position of the bright ribbons.
In (b2), two plasma blobs are marked by green arrows.
An animation of this figure is available, showing the initial phase of the solar jet.
It covers a duration of $\sim$9 minutes from 02:15 to 02:24 UT on 2022 November 12.}
\label{fig:jetfoot}
\end{figure*}

In this study, the EUV wave may be a fast-mode MHD shock wave driven by the ejection of the solar jet.
Figure\,\ref{fig:jetfoot} and its associated animation reveal the evolution of the footpoint region of the solar jet, as depicted in the NVST \ha-center and AIA 211\,\AA\ images.
Figure\,\ref{fig:jetfoot}(a1) and (b1) show images of the NVST \ha-center and AIA 211\,\AA\ passbands prior to the eruption.
At $\sim$02:18:45 UT, bright ribbons appeared inside the footpoint region, indicating a flare occurrence and eventually a small-filament eruption.
During the evolution of this event, the bright ribbons were also visible and expanded, as shown by the red contours in Figure\,\ref{fig:jetfoot}(a2) and (b2).
An untwisting motion typical for jets \citep{2008ApJ...680L..73P,2021ApJ...911...33C} appeared in the NVST \ha-center and AIA 211\,\AA\ images in the associated animation of Figure\,\ref{fig:jetfoot}.
In the meantime, plasma blobs appeared in the fringe of the event (marked by green arrows in Figure\,\ref{fig:jetfoot}(b2)),
  which might be a result of magnetic reconnection between the rising filament and the overlaying loops.
These observational features correspond to the initial phase of the solar jet, which may result from the rise of the small-filament and its interaction with the background magnetic field,
  indicating that the jet is likely a blowout jet \citep[e.g.,][]{2010ApJ...720..757M,2013RAA....13..253H,2015ApJ...815...71C,2016SSRv..201....1R,2017ApJ...851...67S,2022ApJ...927..127S,2023NatCo..14.2107C,2023A&A...673A..83L}.

The solar jet was ejected rapidly about 1.5 minutes after the appearance of the plasma blobs.
This corresponds to the ejection phase of the jet that was followed by the EUV wave, as shown in Figure\,\ref{fig:maps} and \ref{fig:stimg} and the associated animation.
Furthermore, the type II radio burst was detected less than one minute after the start of the EUV wave.
Based on the results from the EUV images and radio dynamic spectrum including the temporal relationship between the blowout jet, EUV wave, and type II radio burst,
  we suggest that the latter two may be the manifestation of the coronal shock driven by the ejection of the blowout jet.

\subsection{Modified velocity of the coronal shock}
\label{subsec:turev}

We have described a coronal shock probably driven by the ejection of a solar jet near the solar disk center.
This shock manifested as an EUV wave captured by the EUV imagers and a type II radio burst recorded by the CBSm radio dynamic spectrum.
The EUV wave's velocity obtained from AIA observations is the projected velocity of the shock,
  while the velocity of the radio burst is the radial one against the coronal density gradient.
Using these two components of the shock velocities, we determined a modified shock velocity to be $\sim$757 \kms.

The frequency band-split during a type II radio burst is generally attributed to the plasma emission from the upstream and downstream shock regions.
By utilizing Equation\,\ref{fun:fn}, we estimated the density ratios across the shock fronts.
The density ratios ($X$) of the shock upstream to downstream were found to be in the range from 1.18 to 1.23,
  which is consistent with those in the previous studies \citep[e.g.,][]{2001A&A...377..321V,2015ApJ...812...52D}, during the time interval of 02:26:00--02:27:30 UT.
Given the following relationships between the Alfv\'en Mach number ($M_{A}$) and $X$ for a parallel shock
\begin{equation}
M_{A} = \sqrt{X},
\label{fun:mapara}
\end{equation}
and for a perpendicular shock
\begin{equation}
M_{A} = \sqrt{\frac{X(X+5)}{2(4-X)}},
\label{fun:maper}
\end{equation}
\citep{1982GAM....21.....P,2002A&A...396..673V}, we have determined the value of $M_{A}$ from the density ratios.
In this study, the value of $M_{A}$ is found to be 1.09--1.11 for a parallel shock geometry and 1.14--1.18 for a perpendicular shock geometry.
Since $M_{A}$ represents the ratio between the shock velocity and the local Alfv\'en velocity, we have inferred that the shock velocity in our study is 10\%--20\% larger than the local Alfv\'en velocity.

\section{Discussion}
\label{sec:dis}

Previous studies have reported that the starting frequencies of type II radio bursts associated with solar jets are in the range of 44--170 MHz, and 
  the source regions of the radio burst are found to be located at a distance of more than 0.5 solar radii above the solar surface \citep{2020ApJ...893..115C,2021ApJ...909....2M,2022ApJ...926L..39D,2023arXiv230511545M}.
In contrast, the type II radio burst observed in this study shows a drift from a starting frequency of 200--250 MHz,
  which is larger than those reported by \citet{2020ApJ...893..115C}, \citet{2021ApJ...909....2M}, \citet{2022ApJ...926L..39D}, and \citet{2023arXiv230511545M}.
On the other hand, \citet{2015ApJ...804...88S} reported that the reconnection between magnetic loops stongly bent toward the solar surface and nearby emerging flux
  could generate a coronal shock at a height of less than 0.1 solar radius above the limb.
Similarly, our findings suggest that the blowout jet occurs in a far-reaching coronal loop system and is not related to a CME.
The radio burst may be generated by the ejection of the blowout jet in the low solar corona and originated at a lower height than CME-driven type II radio bursts.

Our findings show that the type II radio burst is associated with a blowout jet that does not evolve into a CME.
The starting frequency of the radio bursts is in the range of 200--250 MHz, which is larger than those of CME-driven type II radio bursts found in many previous studies
  \citep[e.g.,][]{2007A&A...461.1121C,2008A&A...491..873C,2013ApJ...767...29F,2014ApJ...787...59C,2020FrASS...7...17M,2021A&A...654A..64J,2023ApJ...943...43R}.
This could be the difference between jet-driven type II radio bursts and CME-driven type II radio bursts.

The Alfv\'en Mach number is estimated to be 1.09--1.18 in this study, indicating that the coronal shock may need a continuous driver.
In this case, for a far-reaching coronal loop system, a type II radio burst may involve a frequency decrease pattern and a frequency increase pattern,
  while the latter corresponds to the propagation of the driver after reaching the top of the loop system. 
However, only a persisting frequency decrease pattern is observed during the type II radio burst in this study, which likely contradicts the notion that the blowout jet occurs in a far-reaching coronal loop system.
Two possible scenarios may explain this observational contradiction.
One scenario is that the ejection of the blowout jet can not drive a coronal shock before it reaches the top of the loop system, where the ejection velocity is larger than the local magnetoacoustic velocity.
In this scenario, the frequency of the type II radio burst may continue to decrease for a while and not increase.
The ejected plasma reaches another footpoint of the loops at the end of the ejection ($\sim$02:29:45 UT), which is more than 1 minute after the disappearance of the radio burst ($\sim$02:28:20 UT).
Another scenario is that the ejection of the blowout jet can persistently drive the coronal shock until it reaches the top of the loop system, after which it is not a driver anymore.
Moreover, the disappearance of the radio burst is consistent with the time when the ejection reaches the top of the loop system.
In this case, only the frequency decrease pattern can be observed during the type II radio burst.

Transient and collimated plasma ejections, known as solar jets, are frequently observed along open fields or far-reaching coronal loops \citep[e.g.,][]{2016SSRv..201....1R,2020ApJ...897..113H}.
Small-scale filament eruptions could sometimes generate blowout jets under open field conditions, which evolve into narrow white-light jets or narrow CMEs
  \citep[e.g.,][]{2011ApJ...738L..20H,2019ApJ...881..132D}.
Although type II radio bursts are commonly associated with CMEs \citep[e.g.,][]{2008SoPh..253..215V,2009ApJ...691L.151L,2013ApJ...767...29F}, their relationship with blowout jets has rarely been reported.
\citet{2022ApJ...926L..39D} reported a blowout jet and an associated type II radio burst, which were accompanied by a jet-like CME, but further analysis of their relationship was not conducted.
In our study, we observed an EUV wave and a type II radio burst that were both associated with the ejection of a blowout jet, which did not evolve into a CME.

Small-scale eruptions can drive EUV waves \citep[e.g.,][]{2012ApJ...753L..29Z,2013ApJ...764...70Z,2018ApJ...861..105S,2022ApJ...931..162Z}.
In this study, we observed an EUV wave associated with a blowout jet.
The EUV wave was closely related to the blowout jet and propagated with a similar velocity.
Additionally, a dimming region between the blowout jet and the EUV wave was also observed.
These observations indicate that the EUV wave observed in the AIA 193\,\AA\ and 211\,\AA\ images in this study is probably driven by the ejection of the blowout jet.

Several studies have reported Type II radio bursts in association with solar jets \citep{2015ApJ...804...88S,2020ApJ...893..115C,2021ApJ...909....2M,2022ApJ...926L..39D,2023arXiv230511545M}.
However, they did not provide further details regarding the relationship between the blowout jet and the type II radio burst.
Our study presents more details about the evolution of the blowout jet and its relation to the EUV wave and type II radio burst.
We also found that the evolution of the blowout jet involved an initial phase with a velocity of $\sim$370$\pm$19\,\kms, followed by an ejection phase with a higher velocity of $\sim$560$\pm$87\,\kms.
The EUV wave appeared slightly after the start of ejection phase, and 46\,s before the type II radio burst. 
These findings suggest that both the type II radio burst and the EUV wave are closely linked to the ejection phase of the blowout jet and are likely manifestations of a coronal piston shock driven by the ejection of the blowout jet.

Type II radio bursts are often associated with coronal shocks driven by the propagations or expansions of CMEs from the Sun \citep[e.g.,][]{2011LRSP....8....1C}.
Besides, type II radio bursts might be generated by solar flares \citep[e.g.,][]{2012ApJ...746..152M}.
In this study, we identify a type II radio burst driven by the ejection of a blowout jet.
These indicate that various solar phenomena (CMEs, flares, and jets) can drive type II radio bursts.
What physical conditions (e.g., velocity, $M_{A}$, and 3D magnetic structure) are needed for the generation of type II radio busts is still an open question \citep{2022ApJ...929..175S}.
When the propagation velocity of a jet exceeds the local magnetoacoustic velocity, a shock wave is generated, which in turn produces a type II radio burst.
While, compared to the CMEs and solar flares, type II radio bursts associated with solar jets are relatively rare.
Here, we report a type II radio burst that is probably driven by a solar jet.

\section{Summary}
\label{sec:sum}

In this study, we have presented analysis results of a solar jet accompanied by an EUV wave and a type II radio burst
  with images taken by SDO/AIA, SATech-01/SUTRI, NVST, and STEREO/EUVI and the CSO/CBSm radio dynamic spectrum.
The solar jet is associated with a C4.5 class flare and a small filament eruption, but does not evolve into a CME.
The evolution of the blowout jet involves an initial phase and an ejection phase.
Our findings suggest that the solar jet is a blowout jet occurring in a far-reaching coronal loop system.
The EUV wave and the type II radio burst closely follow the ejection of the blowout jet,
  suggesting that they are likely both manifestations of the coronal shock probably driven by the ejection of the blowout jet.

The AIA and SUTRI EUV images reveal that the solar jet rapidly ejects southward with a velocity of 560$\pm$87\,\kms\ after an initial rising phase with a velocity of 370$\pm$19\,\kms.
Furthermore, an EUV wave propagates southward ahead of the jet with a velocity of 403$\pm$84\,\kms.
In between the head of the solar jet and the EUV wave, a dimming region appears in the AIA 211\,\AA\ and 193\,\AA\ passbands.
The source region of the type II radio burst starts at a lower height than the CME-driven type II radio bursts.
As the EUV wave and the type II radio burst are both the manifestations of the coronal shock in different observations and the driver is close to the solar center,
  the velocity of the EUV wave from the time-distance diagram and the velocity from the radio dynamic spectrum represent the projected velocity of the shock in the plane of sky and its radial velocity, respectively.
Additionally, the radial velocity from the radio dynamic spectrum is estimated to be 641\,\kms.
The combination of the radial velocity of the radio burst and the projected velocity of the EUV wave yields a modified velocity of the shock at 757\,\kms.
The Alfv\'en Mach number is estimated to be 1.09--1.18, implying that the shock velocity is 10\%--20\% larger than the local Alfv\'en velocity.

\acknowledgments
This work was supported by NSFC grant 11825301, National Key R\&D Program of China No. 2022YFF050380 and No. 2020YFC2201200, and China Postdoctoral Science Foundation No. 2021M700246.
MM  acknowledges DFG-grant WI 3211/8-1. This research is partially supported by the Bulgarian National Science Fund, grant No KP-06-N44/2.
AIA and HMI are instruments onboard the Solar Dynamics Observatory, a mission for NASA's Living With a Star program.
SUTRI is a collaborative project conducted by the National Astronomical Observatories of CAS, Peking University, Tongji University, Xi'an Institute of Optics and Precision Mechanics of CAS and the Innovation Academy for Microsatellites of CAS.
The \ha\, data used in this paper were obtained with the New Vacuum Solar Telescope in Fuxian Solar Observatory of Yunnan Astronomical Observatories, CAS.
We thank CSO, GOES and STEREO/SECCHI for providing data.

\bibliographystyle{aasjournal}
\bibliography{bibliography}

\end{document}